\documentclass[aps,prl,amsmath,amssymb,notitlepage,10pt]{revtex4-1}

\usepackage[pdftex]{graphicx}

\usepackage{bm}
\usepackage{color}
\usepackage{fullpage}

\begin{document}

\title{Information-to-work conversion by Maxwell's demon in a superconducting circuit-QED system}

\author{Y.~Masuyama$^{1}$}
\author{K.~Funo$^{2}$}
\author{Y.~Murashita$^{3}$}
\author{A.~Noguchi$^{1}$}
\author{S.~Kono$^{1}$}
\author{Y.~Tabuchi$^{1}$}
\author{R.~Yamazaki$^{1}$}
\author{M.~Ueda$^{3,4}$}
\author{Y.~Nakamura$^{1,4}$}

\affiliation{$^{1}$Research Center for Advanced Science and Technology (RCAST), The University of Tokyo, Meguro-ku, Tokyo 153-8904, Japan}
\affiliation{$^{2}$School of Physics, Peking University, Beijing 100871, China}
\affiliation{$^{3}$Department of Physics, The University of Tokyo, 7-3-1 Hongo, Bunkyo-ku, Tokyo 113-0033, Japan}
\affiliation{$^{4}$Center for Emergnent Matter Science (CEMS), RIKEN, Wako, Saitama 351-0198, Japan}

\date{\today}

\maketitle

%%%%%%%%%%%%%%%%%%%%%%%%%%%%%%%%%%%%%%%%%%%%%%%%%%%%%%%%%%%%%%%%%%%%%%%%%%%%%%%%%%%%%%%%%%%%%%

%\section{Introduction}
\textbf{
The gedanken experiment of Maxwell's demon has led to the studies concerning the foundations of thermodynamics and statistical mechanics~\cite{Leff2002}.
The demon measures fluctuations of a system's observable and converts the information gain into work via feedback control~\cite{Sagawa2008}.
Recent developments have elucidated the relationship between the acquired information and the entropy production and generalized the second law of thermodynamics and the fluctuation theorems~\cite{Parrondo2015,Toyabe2010, Koski2014,Vidrighin2016}.
Here we extend the scope to a system subject to quantum fluctuations by exploiting techniques in superconducting circuit quantum electrodynamics~\cite{Blais2004}. 
We implement Maxwell's demon equipped with coherent control and quantum nondemolition projective measurements on a superconducting qubit, where we verify the generalized integral fluctuation theorems~\cite{Funo2015,Funo2013} and demonstrate the information-to-work conversion.
This reveals the potential of superconducting circuits as a versatile platform for investigating quantum information thermodynamics under feedback control, which is closely linked to quantum error correction~\cite{Terhal2015} for computation~\cite{Kelly2015} and metrology~\cite{Unden2016}.
}\\

The fluctuation theorem is valid in systems far from equilibrium and can be regarded as a generalization of the second law of thermodynamics and the fluctuation-dissipation theorem~
\cite{Jarzynski2010,Campisi2011}.
In particular, the generalized integral fluctuation theorem, which incorporates the information content on equal footing with the entropy production, bridges information theory and statistical mechanics~\cite{Sagawa2010}, and has been extended to quantum systems~\cite{Morikuni2011,Funo2013}.
Experimentally, Maxwell's demons were implemented in classical systems using colloidal particles~\cite{Toyabe2010}, a single electron box~\cite{Koski2014}, and a photodetector~\cite{Vidrighin2016}.
More recently, the integral quantum fluctuation theorem in the absence of feedback control was tested with a trapped ion~\cite{An2014}. 
Maxwell's demon and the generalized second law in a quantum system were studied in spin ensembles with nuclear magnetic resonance~\cite{Camati2016}.
However, experimental demonstrations of the fluctuation theorems that directly address the statistics of single quantum trajectories under feedback control are still elusive.
Toward this goal, recent progress in superconducting quantum circuits offers quantum non-demolition~(QND) projective measurement of a qubit~\cite{Blais2004} and improved coherence times~\cite{Oliver2013} which altogether enable high-fidelity feedback operations. 
For example, stabilization of Rabi oscillations using coherent feedback~\cite{Vijay2012,Campagne-Ibarcq2013}, fast initialization of a qubit~\cite{Riste2012}, and deterministic generation of an entanglement state between two qubits~\cite{riste2013deterministic} have been achieved.

%\section{Experiment}

Here we verify the generalized integral fluctuation theorem under feedback control by using a superconducting transmon qubit as a quantum system and taking statistics over repeated single-shot measurements on individual quantum trajectories.
Note that Nagihloo {\it et al.}~recently reported a related experiment with continuous weak measurement and feedback~\cite{Naghiloo2017}. 
We investigate the role of absolute irreversibility associated with the projective measurements as well~\cite{Funo2015}.

\begin{figure}[tb]
 \begin{center}
    \includegraphics[width=7.8cm]{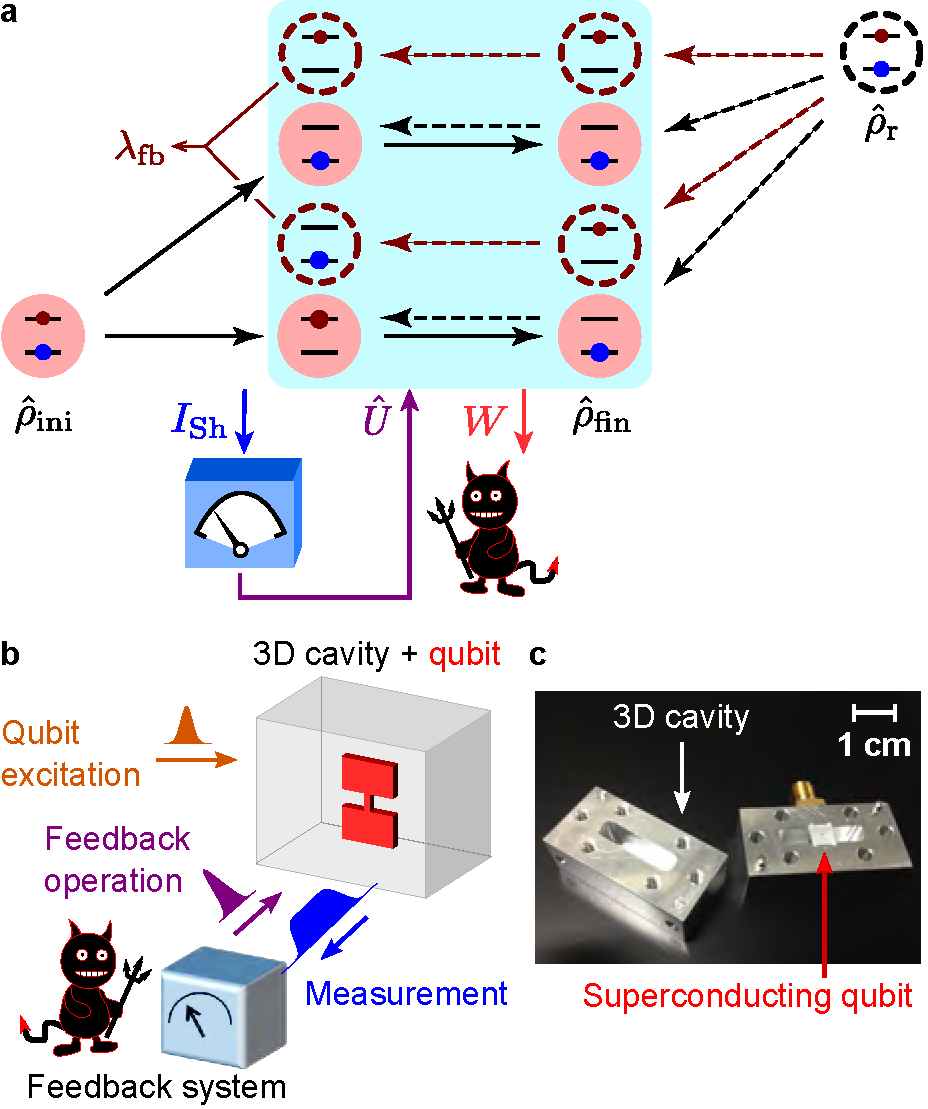}
  \caption{
Maxwell's demon and absolute irreversibility. 
(a)~Concept of the experiment. 
The system initially prepared in a canonical distribution $\hat{\rho}_\mathrm{ini}$ evolves in time. 
A projective measurement by the demon disrupts the evolution, projecting the system onto a quantum state. 
The demon gains the stochastic Shannon entropy $I_\mathrm{Sh}$ and converts it into work $W$ via a feedback operation $\hat{U}$. 
For achieving the ultimate bound of the extracted work $\langle W \rangle =k_\mathrm{B} T \langle I_\mathrm{Sh} \rangle$, the final state distribution $\hat{\rho}_\mathrm{fin}$ of the system has to be the same as $\hat{\rho}_\mathrm{ini}$. 
However, an unoptimized feedback operation prevents it and introduces absolute irreversibility, quantified by the probability $\lambda_\mathrm{fb}$, limiting the amount of the extractable work~[Eq.(\ref{eq:2nd_law})]. 
The time-reversed reference process starts from $\hat{\rho}_\mathrm{r}$~$(= \hat{\rho}_\mathrm{ini})$.
(b)~Schematic of the feedback-controlled system in the experiment. 
(c)~Qubit-resonator coupled system. 
A superconducting transmon qubit fabricated on a sapphire substrate is placed at the center of an aluminum cavity resonator.
In the qubit measurement, the ground and excited states are distinguished in the phase of a microwave readout pulse reflected by the resonator.
}
 \label{fig:(schem)FB_8}
 \end{center}
\end{figure}

The theorem is formulated by considering a pair of processes, the original (forward) process and its time-reversed reference process~[Fig.~\ref{fig:(schem)FB_8}(a)]. 
The initial state of each process is set to be the canonical distribution at temperature $T$.
If we ignore the relaxation of the qubit, the fluctuation theorem reads~\cite{Funo2015,suppl} 
\begin{equation}\label{eq:JEFBabs}
\langle \mathrm{e}^{-\sigma - I_\mathrm{Sh}}  \rangle  = 1 -\lambda_\mathrm{fb},
\end{equation}
where $I_\mathrm{Sh}$ is the stochastic Shannon entropy of the initial state of the qubit, $\sigma = - \beta (W + \Delta F)$ is the entropy production, $\beta$ is the inverse temperature $1/(k_\mathrm{B} T)$ of the qubit, $W$ is the work extracted from the qubit, and $\Delta F$ is the change in the equilibrium free energy of the system.
The constant $\lambda_\mathrm{fb}$ on the right-hand side of Eq.(\ref{eq:JEFBabs}) denotes the total probability of those events in the time-reversed process whose counterparts in the original process do not exist. 
Such events, called absolutely irreversible events, involve a formal divergence of the entropy production and should therefore be treated separately~\cite{Funo2015,suppl}.  
Here, the absolute irreversibility is caused by the projective measurement that restricts possible forward events.
Below, we focus on the case with $\Delta F = 0$, i.e., to the process with the same system Hamiltonian at the beginning and the end, for simplicity of discussions.

\begin{figure*}[tb]
 \begin{center}
    \includegraphics[width=114.4mm]{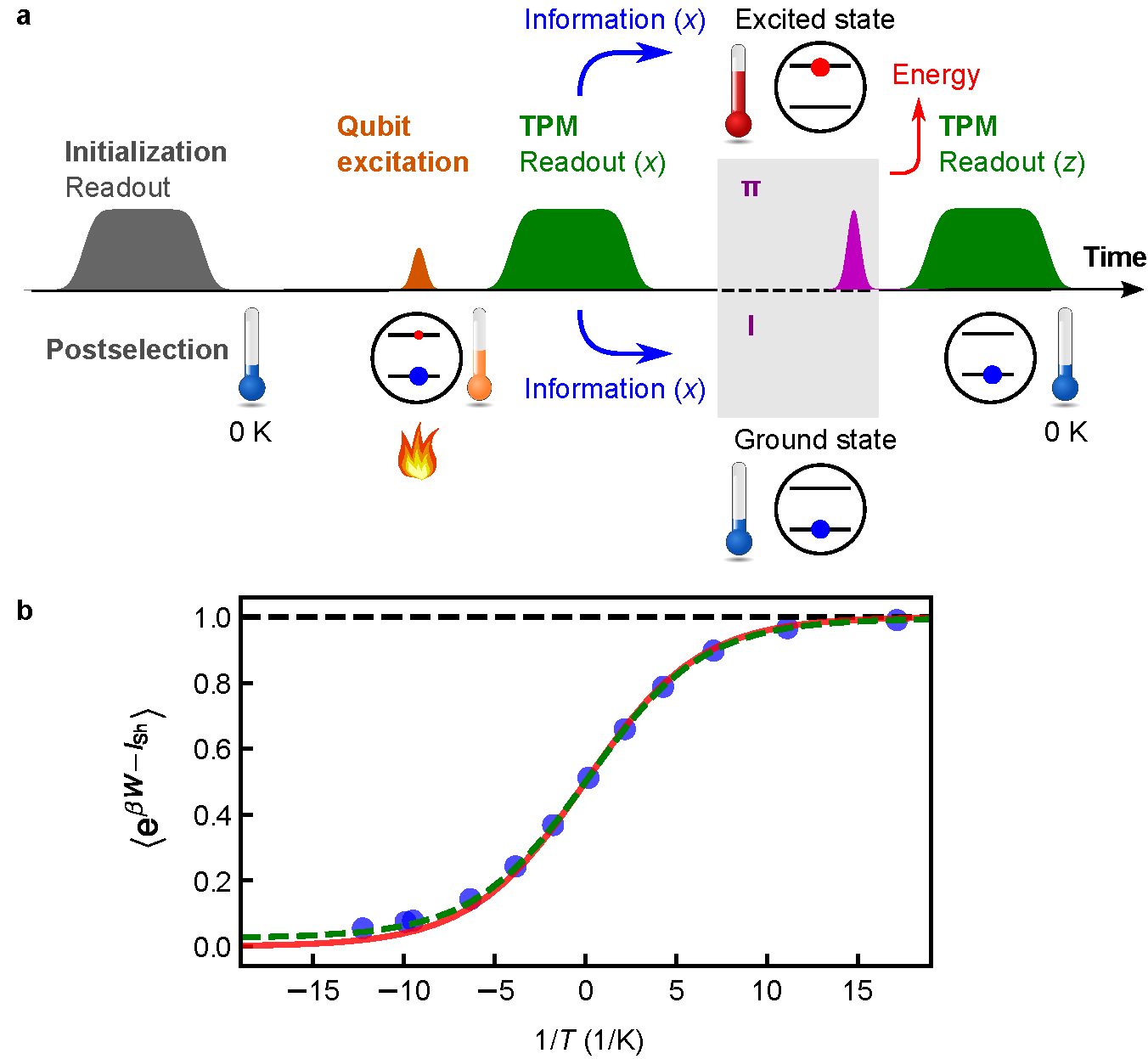}
  \caption{
Generalized integral fluctuation theorem under feedback control.
(a) Pulse sequence used in the experiment. 
The qubit is initialized with a projective measurement and postselection, followed by a resonant pulse  excitation which prepares a superposition as an input. 
The two-point measurement protocol (TPM) consists of two quantum nondemolition projective readout pulses.
Depending on the outcome $x$ of the first readout ($x={\rm g}$ or ${\rm e}$ corresponding to the ground or the excited state of the qubit), a $\pi$-pulse for the feedback control is or is not applied.
The $\pi$-pulse flips the qubit state to the ground state and extracts energy. 
The second readout with outcome $z$ completes the protocol.
See the Supplementary Information~\cite{suppl} for details.
(b)~Experimentally obtained statistical average $\langle \mathrm{e} ^{\beta W - I_\mathrm{Sh}} \rangle$ vs.\ the inverse initial qubit temperature $1/T$ (blue circles). 
The red solid (black dashed) curve is the theoretical value of the probability $1 -\lambda_\mathrm{fb}$ in the presence (absence) of absolute irreversibility.
The green dashed curve is obtained by a master equation taking into account the qubit relaxation during the pulse sequence. 
}
 \label{fig:160508CDK36_649_JEFB_3RO_ctSwp}
 \end{center}
\end{figure*}

In the experiment [Fig.~\ref{fig:160508CDK36_649_JEFB_3RO_ctSwp}(a)], we evaluate the work $W=E(x) - E(z)$ extracted from the system by employing the two-point measurement protocol~(TPM), in which QND projective measurements on the energy eigenbasis (with outcomes $x$ and $z$) are applied respectively to the initial and final states of the system~\cite{Campisi2011}.
A positive amount of the work~($W > 0$) corresponds to the energy deterministically extracted from the system via the stimulated emission of a single photon induced by the $\pi$-pulse. 
Depending on the measurement outcome $x$ for the feedback control, the feedback operation does or does not flip the state of the qubit with a $\pi$-pulse. 
The probability $p(x)$ of the state $x$ being found gives $I_\mathrm{Sh} = -\ln p(x)$.

%\subsection{Dependence on effective tempetature}

In Fig.~\ref{fig:160508CDK36_649_JEFB_3RO_ctSwp}(b) we compare the experimentally obtained statistical average $\langle \mathrm{e} ^{\beta W - I_\mathrm{Sh}} \rangle$ with the theoretical value of $1-\lambda _\mathrm{fb}$~\cite{suppl}.
Depending on the effective temperature of the qubit initial state, the probability of the absolutely irreversible events varies.
The excellent agreement confirms the generalized integral fluctuation theorem under feedback control.
Furthermore, the relation in Eq.(\ref{eq:JEFBabs}) is proven to hold for any initial effective temperature of the qubit, even at negative temperatures.  
The smaller the inverse temperature $\beta$ is, the larger the contribution of absolute irreversibility is.

\begin{figure*}[tb]
 \begin{center}
 \includegraphics[width=158.5mm]{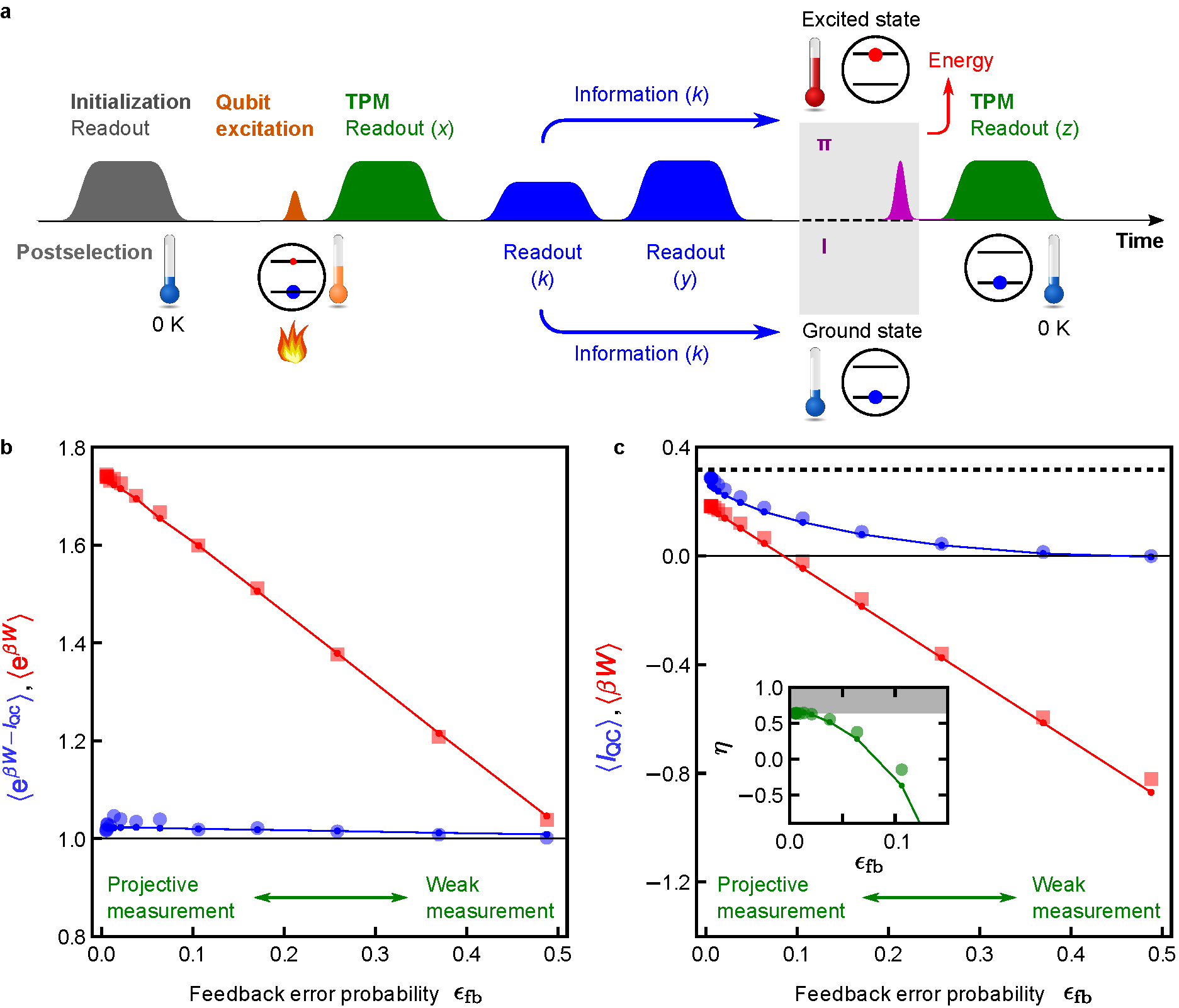}
  \caption{
Effects of the feedback error on the fluctuation theorem and the second law of thermodynamics.
(a)~Pulse sequence. 
Two readout pulses are inserted to the one in Fig.~\ref{fig:160508CDK36_649_JEFB_3RO_ctSwp}(a). 
The outcome $k$(=$\mathrm g$ or $\mathrm e$) obtained by the readout with a variable pulse amplitude is used for the feedback control.
The feedback error probability $\epsilon_{\mathrm{fb}}$ is a function of the measurement strength. 
The subsequent readout with outcome $y$ projects the qubit state before the feedback control.
See Ref.~\cite{suppl} for details.
(b)~Experimentally determined $\langle \mathrm{e}^{\beta W - I_\mathrm{QC}} \rangle$ (blue circles) and $\langle \mathrm{e}^{\beta W} \rangle$ (red squares) vs.\ $\epsilon_{\mathrm{fb}}$.
(c)~$\langle I_\mathrm{QC} \rangle$ (blue circles) and  $\langle  \beta W \rangle$ (red squares) vs.\ $\epsilon_{\mathrm{fb}}$. 
The black dotted line represents the Shannon entropy $\langle I_\mathrm{Sh} \rangle$ of the qubit initial state, which is prepared at the effective temperature $T=0.14$~K with the excited state occupancy of 0.097.
Line-connected dots in (b) and (c) show the simulated results incorporating the effect of qubit relaxation~\cite{suppl}.
Inset in~(c): Information-to-work conversion efficiency $\eta$ (circles) and the simulated result (line-connected dots). 
The efficiency $\eta$ in the gray zone is inaccessible due to the absolute irreversibility. 
}
 \label{fig:160508CDK36_651_JEFB_5RO_roSwp}
 \end{center}
\end{figure*}

%\subsection{Measurement strength dependence}
Next, we investigate the effects of imperfect projection in the readout.
With a weak readout pulse, the state of the qubit is not completely projected.
It also gives less information gain for the feedback control.
To evaluate the influence of the weak measurement, we add two more readout pulses to the pulse sequence  [Fig.~\ref{fig:160508CDK36_651_JEFB_5RO_roSwp}(a)]. 
The TPM again starts with a projective readout with outcome $x$, but now the feedback control is performed based on the subsequent variable-strength measurement with outcome $k$($= \mathrm{g}$ or $\mathrm{e}$).
Then, to project the qubit state before the feedback control, we apply another strong measurement to obtain outcome $y$($= \mathrm{g}$ or $\mathrm{e}$).
Using these measurement outcomes, we calculate the stochastic QC-mutual information $I_\mathrm{QC} = \ln p(y|k) - \ln p(x)$~\cite{Funo2013}.
Here, QC indicates that the measured system is quantum and the measurement output is classical~\cite{Sagawa2008}, and $p(y|k)$ is the probability of outcome $y$ being obtained conditioned on the preceding measurement outcome $k$.
The first term in $I_\mathrm{QC}$ quantifies the correction to $I_\mathrm{Sh}$ because of the imperfect projection.
If the measurement for the feedback control is a QND projective measurement and there is no relaxation of the qubit, $p(y|k)$ becomes unity and $I_\mathrm{QC}$ reduces to $I_\mathrm{Sh}$.
On the other hand, for the measurement with imperfect projection, the absolute irreversibility disappears because such measurement no longer gives restriction on forward events.
Therefore, we obtain $\lambda_\mathrm{fb}=0$.
In this case, the generalized integral fluctuation theorem is reformulated as~\cite{Funo2013,suppl}
\begin{equation}\label{eq:JE5RO}
\langle \mathrm{e}^{\beta W - I_\mathrm{QC}}  \rangle  = 1.
\end{equation}

Figure \ref{fig:160508CDK36_651_JEFB_5RO_roSwp}(b) plots the statistical averages, $\langle \mathrm{e} ^{\beta W - I_\mathrm{QC}} \rangle$ and $\langle \mathrm{e} ^{\beta W} \rangle$, evaluated from the measurement outcomes of the pulse sequence shown in Fig.~\ref{fig:160508CDK36_651_JEFB_5RO_roSwp}(a).
By changing the amplitude of the readout pulse measuring $k$, it is possible to continuously vary the post-measurement state from the projected state to a weakly disturbed state.
Accordingly, the feedback error probability $\epsilon_{\mathrm{fb}}$ increases with decreasing the 
readout pulse amplitude.
(See the Supplementary Information~\cite{suppl} for the details.) 
We see that $\langle {\mathrm{e}}^{\beta W -I_{\mathrm{QC}}} \rangle $ (blue circles), which involves the information gain due to the measurement, is almost unity regardless of the feedback error probability.
The small deviation from unity is understood as the effect of the qubit relaxation during the TPM (blue curve)~\cite{Pekola2015}. 
In contrast, the value $\langle \mathrm{e}^{\beta W} \rangle$, which discards the information used in the feedback operation, clearly deviates from unity. 
For the weaker readout amplitude, however, the amount of information gain becomes less, and thus $\langle \mathrm{e}^{\beta W} \rangle$ approaches unity.
The situation corresponds to the integral fluctuation theorem in the absence of feedback control.

%\section{conversion efficiency}
Figure~\ref{fig:160508CDK36_651_JEFB_5RO_roSwp}(c) depicts the statistical averages $\langle I_\mathrm{QC} \rangle$ and $\langle  \beta W \rangle$ as a function of the feedback error probability $\epsilon_{\mathrm{fb}}$.
The QC-mutual information $\langle I_\mathrm{QC} \rangle$ (blue circles) decreases to zero with increasing $\epsilon_{\mathrm{fb}}$.
Even for $\epsilon_{\mathrm{fb}}=0$, there remains a difference between $\langle I_\mathrm{QC} \rangle$ and $\langle I_\mathrm{Sh} \rangle$ (black dotted line) due to the qubit relaxation between the two readouts for $k$ and $y$.
The difference between $\langle I_\mathrm{QC} \rangle$ and $\langle  \beta W \rangle$ in the limit of $\epsilon_{\mathrm{fb}} \rightarrow 0$ corresponds to $\ln(1- \lambda _\mathrm{fb})$ in this feedback protocol.

The conversion efficiency from the QC-mutual information $\langle I_\mathrm{QC} \rangle$ to the work $\langle W \rangle$ is defined as 
\begin{equation}
\eta=\frac{ \langle W \rangle}{k _\mathrm{B} T \langle I _\mathrm{QC} \rangle},
\end{equation}
where we omit the contribution from the free-energy change by considering $\Delta F = 0$.
As shown in the inset of Fig.~\ref{fig:160508CDK36_651_JEFB_5RO_roSwp}(c), $\eta$ becomes larger for stronger measurement and reaches the maximum value of 0.65. 
The main limiting factor of the efficiency in the present experiment is the contribution
$k_\mathrm{B}T \ln (1 - \lambda_\mathrm{fb})$ in the generalized second law of thermodynamics~\cite{suppl}
\begin{equation}
\langle W \rangle \leq  k_\mathrm{B} T \langle I _\mathrm{Sh} \rangle  + k_\mathrm{B} T \ln (1 -\lambda_\mathrm{fb})
\label{eq:2nd_law}
\end{equation}
which is derived from the fluctuation theorem Eq.(\ref{eq:JEFBabs}).
The result in the inset of Fig.~\ref{fig:160508CDK36_651_JEFB_5RO_roSwp}(c) indicates that our feedback scheme achieves the equality condition in Eq.(\ref{eq:2nd_law}) and is optimal in this sense.

%\section{summary}

We have successfully implemented Maxwell's demon and verified the generalized integral fluctuation theorem 
in a single qubit.
In the present work, the measurement outcome obtained by the demon was analyzed in terms of the Shannon and the QC-mutual information.
On the other hand, the effect of the coherence can be investigated in a similar setup~\cite{Elouard2017}.
By implementing the memory of the demon with a qubit~\cite{Quan2006}, or a quantum resonator as demonstrated recently~\cite{Cottet2017}, one can characterize the energy cost for the measurement~\cite{Sagawa2009} or study feedback schemes maintaining the coherence between the system and the memory to improve the energy efficiency of the feedback.
Superconducting quantum circuits further allow us to extend the study of information thermodynamics to larger and more complex quantum systems.
It will lead to an estimation of the lower bound of the thermodynamic cost for quantum information processing.

\section*{Methods}
The transmon qubit has the resonant frequency $\omega_\mathrm{q}/2\pi = 6.6296$~GHz, the energy relaxation time $T_1=24$~$\mu$s, and the phase relaxation time $T_2^\ast = 16$~$\mu$s at the base temperature $\sim$10~mK of a dilution refrigerator. 
The cavity has the resonant frequency $\omega _\mathrm{cav}/2\pi = 10.6180$~GHz, largely detuned from the qubit, and the relaxation time $1/\kappa = 0.076$~$\mu$s. 
The coupling strength between the qubit and the resonator is estimated to be $g/2\pi = 0.14$~GHz.

The pulse sequences for the experiments in Figs.~\ref{fig:160508CDK36_649_JEFB_3RO_ctSwp} and~\ref{fig:160508CDK36_651_JEFB_5RO_roSwp} take about 2.5~$\mu$s and 4~$\mu$s, respectively.
Each readout pulse has the width of 500~ns.
The qubit excitation pulse and the feedback control pulse are both 20-ns wide. 
See~\cite{suppl} for details.
We take the statistics of the outcomes by repeating the pulse sequence about $8\times 10^4$ times, with a repetition interval 300~$\mu$s which is much longer than the qubit relaxation time.

\section*{Acknowledgements}

The authors acknowledge T. Sagawa for useful discussions
and W. D. Oliver for providing the transmon qubit. 
This work was partly supported by JSPS KAKENHI (Grant No.~26220601), NICT, and JST ERATO (Grant No.~JPMJER1601).
Y.Mu.\ was supported by JSPS through the Program for leading Graduate School (MERIT) and JSPS Fellowship (Grant No.~JP15J00410).
K.F.\ acknowledges supports from the National Science Foundation of China (grants~11375012, 11534002).

\section*{Author contributions}

Y.Ma., K.F.\ and Y.Mu.\ designed the experiments. 
Y.Ma.\ conducted the experiments.
S.K.\ and Y.T.\ assisted in setting up the measurement system.
K.F., Y.Mu.\ and M.U.\ provided theoretical supports.
A.N.\, Y.T.\ and R.Y.\ participated in discussions on the analysis. 
Y.Ma.\ and Y.N.\ wrote the manuscript with feedback from all authors.
M.U.\ and Y.N.\ supervised the project.

%%%%%%%%%%%%%%%

%\bibliography{JE_FB}

%%%%%%%%%%%%%%%%%%%%%%%%%%%%%%%%%%%%%%%%%

\newpage

\section{Supplementary Information for ``Information-to-work conversion by Maxwell's demon in a superconducting circuit-QED system''}

\section{Generalized integral fluctuation theorem}

\begin{figure}[b]
 \begin{center}
    \includegraphics[width=16.0cm]{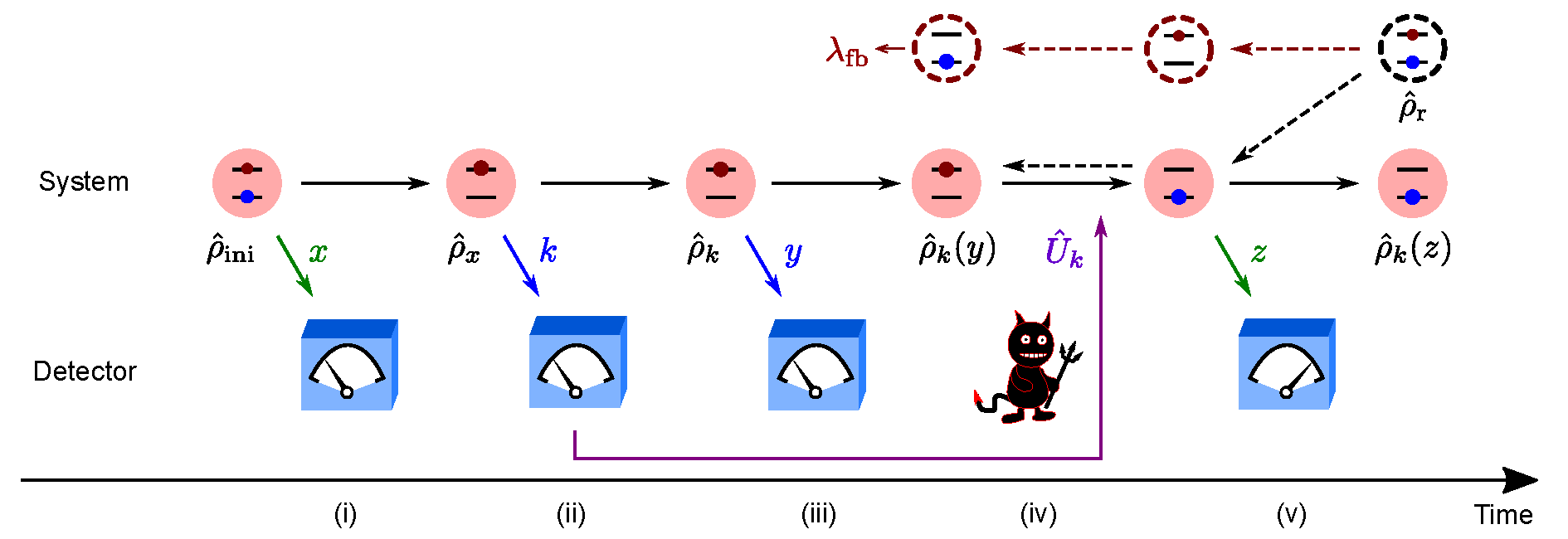}
  \caption{
Schematic illustration of a feedback process of a qubit. 
(i) Projective measurement of the initial state $\hat{\rho}_\mathrm{ini}$. 
Here, for simplicity, we illustrate the case where the initial state $x$ of the qubit is found to be in the excited state. 
(ii) Measurement for the feedback control. The measurement outcome $k$ is recorded in the detector, and a feedback operation [i.e., (iv)] is later performed based on the outcome. 
(iii) Projective measurement of the state $\hat{\rho}_k$ right after the measurement for the feedback control. When the measurement in (ii) is a quantum non-demolition projective measurement on the qubit eigenbasis, the post-measurement state $y$ should be the excited state, and therefore the probability of being found in the ground state (dashed circle) vanishes.  
(iv) Feedback operation based on the outcome $k$. 
As $k$ indicates that the state is in the excited state, a $\pi$-pulse is applied to flip the qubit. 
(v) Measurement of the final state. 
The final state $z$ is found to be in the ground state. 
In the time-reversed process from the reference state $\hat{\rho}_\mathrm{r}$ in the canonical distribution, the system may evolve into the ground state as indicated by the dashed arrow.
This event has no counterpart in the forward process, and therefore contributes to $\lambda_\mathrm{fb}$ which is calculated as the total probability of such transitions.
  }
 \label{fig:(schem)JEFuno}
 \end{center}
\end{figure}

We derive the generalized integral fluctuation theorem under feedback control~\cite{Funo2015} in the context of our experimental protocol.
Figure~\ref{fig:(schem)JEFuno} illustrates an example of a feedback process for a qubit.
We perform a quantum-non-demolition~(QND) projective measurement on the initial canonical state of the qubit
\begin{align}
\hat{\rho}_\mathrm{ini} =& \sum_{x=\mathrm{g,e}} p_{\rm can} (x)| x \rangle \langle x|,
\end{align}
and obtain the state $\hat{\rho}_x= | x \rangle \langle x|$ characterized by the measurement outcome $x$ with probability $p_{\mathrm{can}}(x)$ [Fig.~\ref{fig:(schem)JEFuno}(i)].
Here $| \mathrm{g} \rangle$ and $| \mathrm{e} \rangle$
are the qubit energy eigenstates.
Next, we perform a measurement (outcome $k$) for the feedback control [Fig.~\ref{fig:(schem)JEFuno}(ii)].
Due to the backaction of this measurement, the post-measurement state becomes
\begin{align}
\hat{\rho}_k= \frac{\hat{M}_k \hat{\rho}_{x} \hat{M}^\dagger_k}{p(k)},
\label{eq:JEFunoPostMeasState}
\end{align}
where 
$\hat{M}_k$ and $p(k)$ denote the Kraus operator describing the measurement process and the probability for the outcome~$k$, respectively.
By the subsequent projective measurement (outcome $y$) [Fig.~\ref{fig:(schem)JEFuno}(iii)], one obtains the stochastic QC-mutual information 
\begin{align}\label{eq:stochasticIqc}
I_\mathrm{QC} (x,k,y) = \ln p(y|k) - \ln p_{\rm can} (x),
\end{align}
where $p(y|k)$ is the probability of the outcome $y$ conditioned on the outcome $k$.  
We perform a feedback operation $\hat{U}_k$, which depends on the measurement outcome $ k $ for the feedback control, on the state $\hat{\rho}_k(y) := |y \rangle \langle y|$~[Fig.~\ref{fig:(schem)JEFuno}(iv)]. 
Subsequently, we perform a projective measurement~(outcome $ z $) and obtain the final state $\hat{\rho}_k (z) =  | z \rangle \langle z |$~[Fig.~\ref{fig:(schem)JEFuno}(v)]. 
The probability distribution of the forward process is
\begin{align}
p(x,k,y,z) = p_{\rm can} (x) \, p(k,y\,|\,x) \, p(z\,|\,k,y),
\end{align}
where $p(k,y\,|\,x)=|\langle y |\hat{M}_{k}| x \rangle|^{2}$ and $p(z\,|\,k,y) = | \langle z | \hat{U}_k | y \rangle |^2 $.

Next we introduce the time-reversed process to derive the fluctuation theorem for the dissipated work $ \sigma = - \beta (W + \Delta F)$, where $ W $ denotes the extracted work and $ \Delta F $ is the free-energy difference.
In the present situation, since there is no difference between the initial and final Hamiltonians, $ \Delta F $ vanishes.
The initial state of the time-reversed process, which is called the reference state $\hat{\rho}_\mathrm{r}$, is chosen to be the canonical distribution 
\begin{align}
\hat{\rho}_\mathrm{r}  =&  \sum _{z=\mathrm{g,e}} {p}_{\rm can} (z) | z \rangle \langle z | .
\end{align}
We define the probability distribution under the time-reversed feedback operation as $\tilde{p} (k,y,z) = \tilde p(y\,|\,k,z)\, p(k)\, p_{\rm can}(z) $, where we define the time-reversed transition probability as $\tilde{p}(y \,|\, k,z):=|\langle y |\hat{U}_{k}^{\dagger}| z \rangle|^{2}=p(z\,|\,k,y)$.
The fluctuation theorem can be derived by forming the ratio of the time-reversed probability $\tilde p(k,y,z)$ to the forward probability $p(k,y,z):=\sum_x p(x,k,y,z)$.
To do so, we should separate the set $Y=\{ y\,|\,p(y|k)\neq 0 \}$ to guarantee that the denominator of the ratio does not vanish.
Then, the fluctuation theorem is derived as follows:
\begin{align}
1 
=& \sum_{k,y \in Y,z} \tilde{p} (k,y,z) + \sum_{k,y \notin Y,z} \tilde{p} (k,y,z) \nonumber\\
=& \sum_{x,k,y \in Y,z}p(x,k,y,z)\,  \frac{\tilde{p} (k,y,z)}{p(k,y,z)} + \sum_{k,y \notin Y,z} \tilde{p} (k,y,z)  \nonumber \\
=& \sum_{x,k,y \in Y,z}p(x,k,y,z) \, \mathrm{e}^{\beta W(x,k,z) - I _\mathrm{QC} (x,k,y)} + \sum_{k,y \notin Y,z} \tilde{p} (k,y,z)   \nonumber \\
=& \, \langle \mathrm{e}^{\beta W(x,k,z) - I _\mathrm{QC} (x,k,y)}  \rangle + \sum_{k,y \notin Y,z} \tilde{p} (k,y,z),
\label{eq:derivation_of_FT}
\end{align}
where we use Eq.(\ref{eq:stochasticIqc}) and $-\beta W=\sigma= \ln [p_{\rm can}(x)/p_{\rm can}(z)]  $ to obtain the third line.
For the events with $y\notin Y$, the entropy production formally diverges due to the detailed fluctuation theorem~\cite{Tasaki2000,Funo2013} as $\ln(0/p) = - \infty$.
These events with divergent entropy production (red dotted line in Fig.~\ref{fig:(schem)JEFuno}) are called absolutely irreversible events~\cite{Funo2015}.
To circumvent the problem of divergence, one should subtract the total probability of the absolutely irreversible events defined by
\begin{align}
\lambda_\mathrm{fb} :=\sum_{k,y \notin Y,z} \tilde{p} (k,y,z).
\label{eq:lambda_fb}
\end{align}
As a result, we obtain the generalized integral fluctuation theorem under feedback control in the presence of absolute irreversibility~\cite{Funo2015}:
\begin{align}\label{eq:JE_AbsoIrrev}
\langle \mathrm{e}^{\beta W - I _\mathrm{QC}}  \rangle  = 1 -\lambda_\mathrm{fb}.
\end{align}
By applying Jensen's inequality, we obtain
\begin{align}
\beta \langle W \rangle \leq   \langle I _\mathrm{QC} \rangle  +  \ln (1 -\lambda_\mathrm{fb}).
\label{eq:generalThermoSecond}
\end{align}
The inequality~(\ref{eq:generalThermoSecond}) can be regarded as the generalized second law of thermodynamics that incorporates the effects of feedback control.

If the readout $k$ for the feedback control is not a projective readout, $\lambda_\mathrm{fb}$ vanishes and Eq.(\ref{eq:JE_AbsoIrrev}) reduces to 
$ \langle \mathrm{e} ^{\beta W - I _\mathrm{QC}} \rangle = 1$ as demonstrated in Fig.~3(b).
On the other hand, $I_\mathrm{QC}$ reduces to the stochastic Shannon entropy~$I _\mathrm{Sh}$ when the measurement is a projective measurement as in the experiment shown in Fig.~2.
In this case, the left-hand side of the generalized integral fluctuation theorem can be calculated from the experimentally accessible values as
\begin{align}
 \langle \mathrm{e}^{\beta W -I_\mathrm{Sh}}\rangle 
=& \, \sum_{x,z} p(x,z)\, \mathrm{e}^{\beta W(x,z) - I_\mathrm{Sh} (x)}  \nonumber \\ 
=& \, p(x=\mathrm{g})\, p (z=\mathrm{e} \,|\,x= \mathrm{g} )\, \mathrm{e}^{-\beta \hbar \omega_\mathrm{q} -I(x=\mathrm{g})}
+ p(x=\mathrm{e})\, p (z=\mathrm{g} \,|\, x=\mathrm{e} )\, \mathrm{e}^{\beta \hbar \omega_\mathrm{q}-I(x=\mathrm{e})} \nonumber \\
&+ p(x=\mathrm{e})\, p (z=\mathrm{e} \,|\,x=\mathrm{e} )\, \mathrm{e}^{-I(x=\mathrm{e})} 
+ p(x=\mathrm{g})\, p (z=\mathrm{g} \,|\,x=\mathrm{g} )\, \mathrm{e}^{-I(x=\mathrm{g})},
\label{eq:JEFBexp_lhs}
\end{align}
where $p(x)$ ($x = \mathrm{g, e}$) is the probability of observing the state $x$ in the first measurement of the two-point measurement protocol, the inverse temperature $\beta = (\hbar \omega_{\mathrm{q}})^{-1} \ln \left[p(x=\mathrm{g})/p(x=\mathrm{e})\right]$, and $p(z\,|\,x)$ is the conditional probability of observing the state $z$ ($ z = \mathrm {g, e} $) in the second measurement after observing the state $ x $ in the first one.
Moreover, $I(x)= -\ln p(x) $ denotes the stochastic Shannon information obtained when the state $x$ is observed.
In this case, the maximum value of the extracted work is determined by $\beta \langle W \rangle \leq  \langle I _\mathrm{Sh} \rangle  +  \ln (1 -\lambda_\mathrm{fb})$.
When the equality is achieved, the maximum feedback efficiency with this feedback protocol is also achieved~[see the inset in Fig.~3(c)].

\section{Experimental details}

\subsection{Cryogenic environment}

\begin{figure}[tb]
 \begin{center}
    \includegraphics[width=14.0cm]{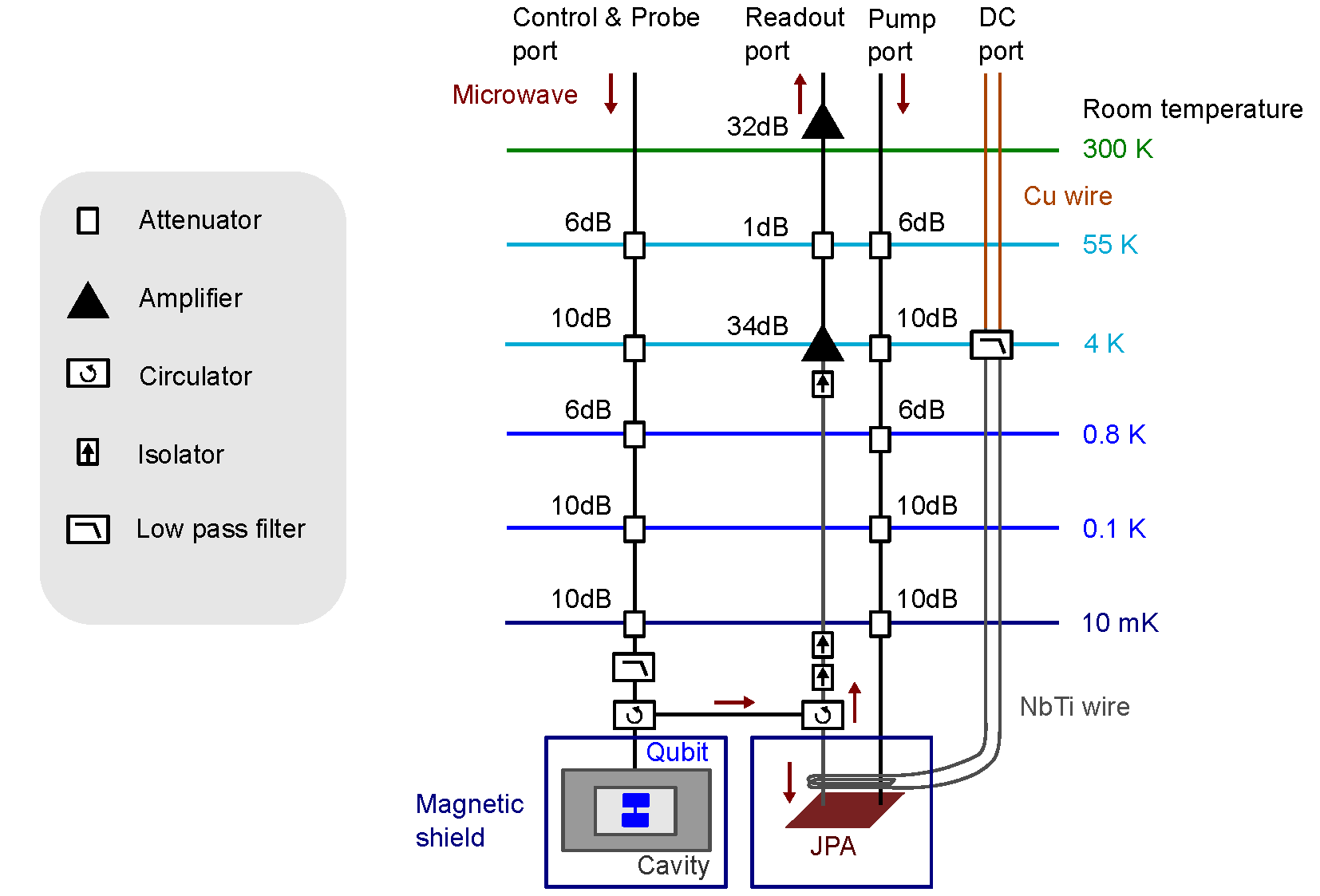}
  \caption{
Schematic of the wiring in the dilution refrigerator.
}
 \label{fig:fridgeLine}
 \end{center}
\end{figure}

The experiment was conducted in a cryogen-free dilution refrigerator with the base temperature of about 10~mK.
The cavity enclosing the supeconducting qubit is placed inside a magnetic shield. 
A flux-driven Josephson parametric amplifier (JPA)~\cite{Yamamoto2008} is placed in a separated magnetic shield and is biased with a small solenoid at a static magnetic field as well as pumped at twice of the resonance frequency.
The probe pulse for the readout is introduced to the cavity through a series of attenuators, and the reflected signal is amplified by the JPA operated in the degenerate mode and by the following amplifiers at 4-K and 300-K stages~(Fig.~\ref{fig:fridgeLine}).

\subsection{Sample}

The circuit quantum-electrodynamical system is constructed with a transmon-type superconducting qubit~\cite{Koch2007} in a cavity.

The transmon qubit is fabricated on a sapphire substrate.
The size of the two aluminum pads of the transmon is $250 \times 500$~$\mu$m$^2$ each, and the area of the Josephson junction is $ 150 \times 250$~nm$^2$.
The qubit has the bare resonance frequency $\omega_\mathrm{q}/2\pi = 6.6296$~GHz and the anharmonicity of   $-345$~MHz. 
The latter is defined as the difference between the excitation frequencies from the ground to the first excited states and from the first to the second excited states. 
From independent time-domain experiments, we obtained the energy relaxation time $T_1=24$~$\mu$s, and the phase relaxation time $T_2^\ast = 16$~$\mu$s. 

The aluminum-made rectangular cavity has the fundamental mode TE$_{101}$ with the bare resonance frequency of $\omega _\mathrm{cav}/2\pi = 10.6180$~GHz, largely detuned from the qubit.
It has a single SMA-connector port, and the relaxation time $1/\kappa = 0.076$~$\mu$s is determined by the sum $\kappa$ of the external and internal loss rates, $\kappa_{\mathrm{ex}}/2\pi = 1.47$~MHz and $\kappa_{\mathrm{in}}/2\pi = 0.63$~MHz, with the qubit mounted inside.

The qubit is mounted at the center of the cavity.
The coupling strength between the qubit and the cavity mode is estimated to be $g/2\pi = 0.14$~GHz from the measurement of the dispersive shift $\chi/2\pi=(g^2/\Delta)/2\pi=-4.6$~MHz of the cavity mode, where $\Delta = \omega_{\mathrm{cav}}-\omega_{\mathrm{q}}$ is the detuning between the qubit and the cavity.
Due to the interaction, the cavity frequency is shifted to $(\omega_{\mathrm{cav}}+g^2/\Delta)/2\pi = 10.6226$~GHz, and the qubit frequency is shifted to $(\omega_{\mathrm{q}}-g^2/\Delta)/2\pi = 6.6342$~GHz.

\subsection{Quantum non-demolition projective readout of the qubit}

In the dispersive regime where the detuning $ \Delta $  is much larger than the qubit-cavity coupling strength $g$, the Jaynes-Cummings Hamiltonian of the coupled system can be approximated as~\cite{Blais2004}
\begin{align}
H_{\mathrm{JC}} &= \hbar \omega_\mathrm{cav} \left( \hat{a}^ \dagger \hat{a} + \frac{1}{2} \right) 
+ \frac{\hbar \omega_\mathrm{q}}{2} \hat{\sigma}_z 
+  \hbar g \left(\hat{a}^ \dagger \hat{\sigma}_- + \hat{a} \hat{\sigma}_+ \right)  \nonumber \\ 
&\approx \hbar \omega_\mathrm{cav} \left( \hat{a}^ \dagger \hat{a} + \frac{1}{2} \right) + \frac{\hbar}{2} \left( \omega_\mathrm{q} - \frac{g^2}{\Delta} \right) \hat{\sigma}_z -  \hbar \frac{g^2}{\Delta} \hat{\sigma} _z \hat{a}^ \dagger \hat{a},
\label{eq:dispersiveH}
\end{align}
where $\hat{a}$ is the annihilation operator of the cavity mode, $\hat{\sigma}_z$ is the Pauli operator  of the qubit, and $\hat{\sigma}_+$ and $\hat{\sigma}_-$ are the qubit raising and lowering operators, respectively.
It indicates that the cavity resonance frequency depends on the states of the qubit, giving rise to the so-called dispersive shift.
Thus, the states of the qubit can be projected onto the energy eignenstates by the measurement of a phase shift of a resonant microwave pulse reflected by the cavity~\cite{Blais2004}.

As the interaction term, i.e., the last term in Eq.(\ref{eq:dispersiveH}), commutes with the qubit Hamiltonian $\propto \hat{\sigma}_z$, the dispersive readout has a quantum non-demolition nature, which is crucial in the present work.

\subsection{Experimental setup for the qubit readout}
We generate the microwave probe pulses for qubit readout by using single-side-band modulation of the continuous carrier microwaves with a 50-MHz intermediate-frequency (IF) signal from a digital-to-analog converter~(DAC; UCSB GHzDAC, 1~GSa/s, 12-bit resolution).
The temporal shape of the pulse, i.e., the amplitude and the phase, is defined by multiplying the waveform from the DAC and the carrier microwaves at an IQ mixer.
We adjust the the microwave carrier frequency of the readout pulses at 10.6219~GHz to obtain the largest difference in the complex amplitude of the electric field between the reflection signals corresponding to the ground state and the excited state of the qubit.

The readout signal, reflected by the cavity and amplified in the chain of the amplifiers in Fig.~\ref{fig:fridgeLine}, passes through a frequency-tunable resonator-type bandpass filter (bandwidth 50~MHz) and is down-converted to the intermediate frequency of 50~MHz with an IQ mixer and a local oscillator~(Fig.~\ref{fig:(schem)FBcircuit}). 
One of the IF ports of the IQ mixer is connected to an analog-to-digital converter (ADC; Acqiris AP240, 1~GSa/s, 8-bit resolution) for signal acquisitions in the readouts $x$, $y$ and $z$ (not shown in the figure), and the other is connected to the feedback system described below.

\begin{figure}[tb]
 \begin{center}
    \includegraphics[width=8.0cm]{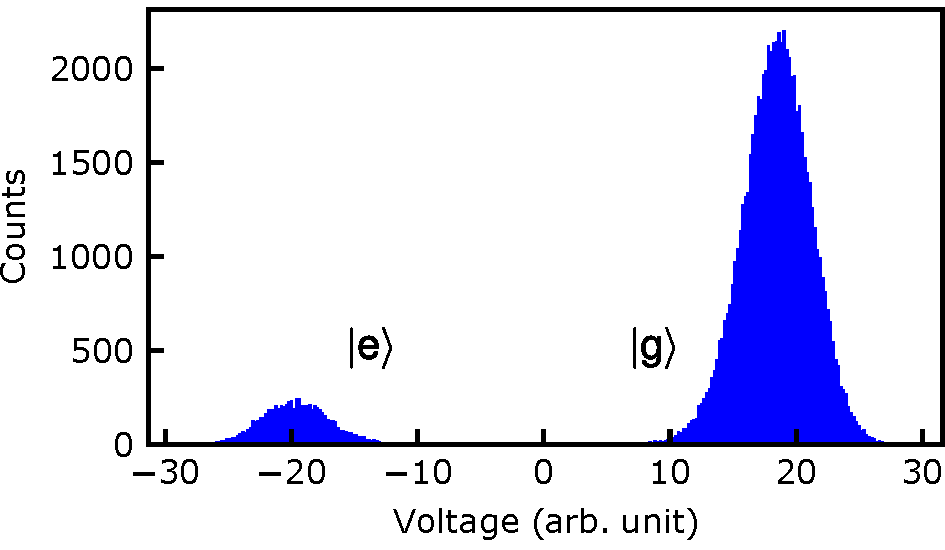}
  \caption{An example of the histogram of the readout outcomes.}
 \label{fig:readout_histogram}
 \end{center}
\end{figure}

The signal acquired in the ADC is digitally processed to discriminate the ground and the excited states. 
The separation of the signals for the 500-ns-wide projective readout pulse is almost 100\%~(Fig.~\ref{fig:readout_histogram}).
Moreover, we checked the quantum non-demolition property of the measurement by two subsequent measurements of the qubit. 
In the test with two 1-$\mu$s-wide readout pulses, 99.6\% of the ground state observed in the first measurement remained in the ground state at the second measurement, and 96.6\% of the excited state stayed the same state.  
The amount of the reductions can be attributed to the energy relaxation of the qubit during the interval (about 0.3~$\mu$s) between the pulses.

\subsection{Experimental setup for the feedback control}
\begin{figure}[tb]
 \begin{center}
    \includegraphics[width=16.0cm]{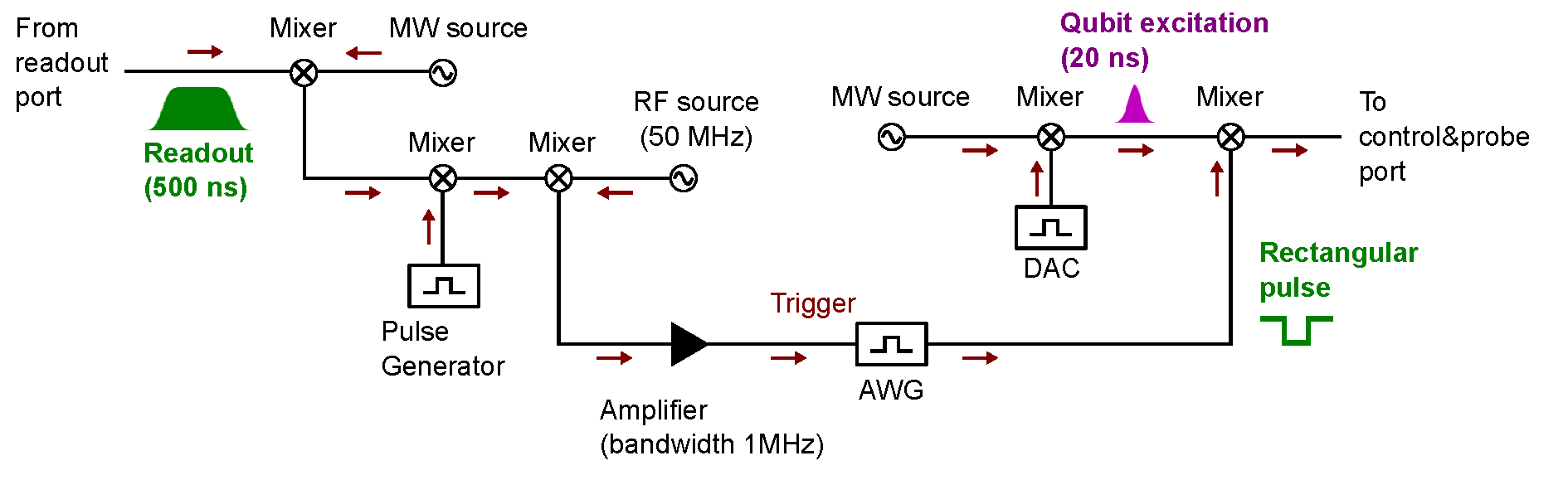}
  \caption{Experimental setup for the feedback control.}
 \label{fig:(schem)FBcircuit}
 \end{center}
\end{figure}

Figure~\ref{fig:(schem)FBcircuit} illustrates the feedback system used in the experiment. 
A $\pi$-pulse is applied to the qubit only if the readout for the feedback control finds the qubit in the excited state.

As shown in Fig.~\ref{fig:(schem)FBcircuit}, the down-converted readout signal at the IF frequency is first chopped at a mixer with a rectangular pulse from a pulse generator (Stanford Research Systems DG535) to select the time window used in the following analog processing, and is further down-converted to zero frequency. 
The obtained pulse signal is amplified with an amplifier (NF Corporation N5307, gain $\times$20, 1-MHz filter bandwidth), and is input to the trigger port of an arbitrary waveform generator (AWG; Tektronix AWG430). 
The threshold of the trigger is adjusted such that the AWG outputs a negative pulse on top of the positive offset voltage only for the input signal corresponding to the qubit ground state.
The triggered~(untriggered) events are recorded as the outcome $k=\mathrm{g}$ ($k=\mathrm{e}$).

The $\pi$-pulse for the feedback control is generated by single-sideband modulation of the continuous microwaves at the qubit drive frequency with a DAC. 
Finally, the voltage pulse from the AWG suppresses the output of the $\pi$-pulse when it is triggered. 
The total delay in the feedback control is about 200~ns measured from the end of the readout pulse.

\subsection{Details of the pulse sequences}

\begin{figure}[tb]
 \begin{center}
    \includegraphics[width=16.0cm]{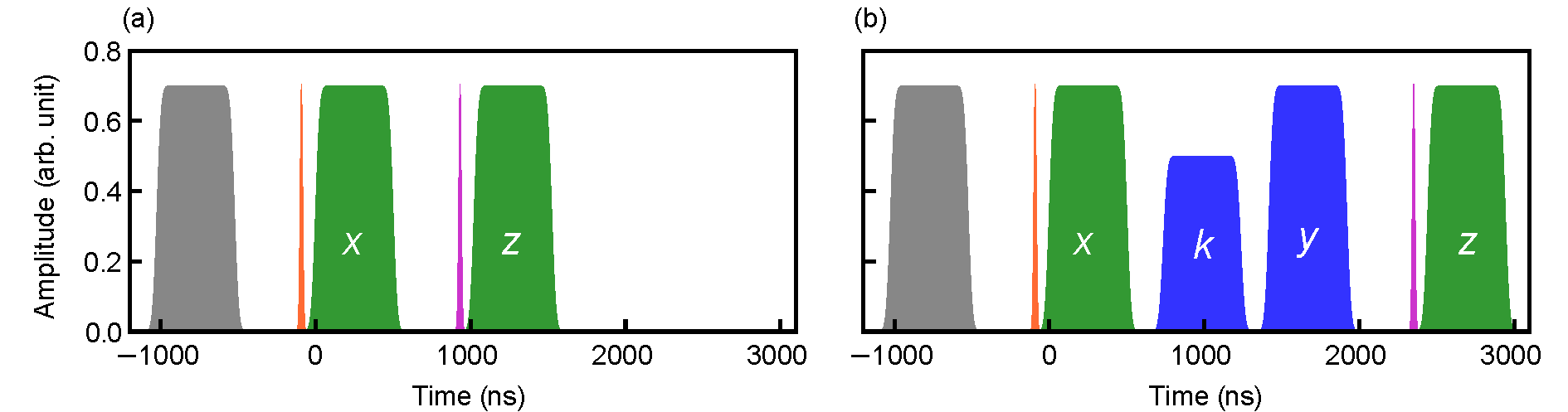}
\caption{
Details of the pulse sequences. 
The origin of the time is set at the beginning of the two-point measurement protocol. 
(a)~Pulse sequence used in the experiment in Fig.~2.
(b)~Pulse sequence used in the experiment of Fig.~3.
The color codes and labels ($x,y,z,k$) are: readout pulse for initialization (gray), qubit excitation pulse (orange), projective readout pulses for the two-point measurement protocol ($ x, z $), $\pi$-pulse for feedback control (magenda), and variable-amplitude readout pulse ($ k $) and subsequent projective readout pulse ($ y $) (blue).
}
 \label{fig:(schem)plsSeq}
 \end{center}
\end{figure}
Figure~\ref{fig:(schem)plsSeq} illustrates the detailed timings of the pulse sequences.
The qubit readout pulses are 500-ns wide with Gaussian-shaped rise and fall edges of 52-ns wide.
The inverse temperature of the qubit, $ \beta $, is determined by the first readout in the two-point measurement protocol by assuming the Boltzmann distribution.
In the absence of the initialization readout and the excitation pulse, the effective temperature of the qubit equilibrated with the cavity field is found to be about 0.16~K, which is significantly higher than the fridge temperature presumably because of the residual noise introduced through the microwave cables.

The qubit control pulse has a Gaussian shape with the width of 20~ns and is applied after the preceding readout pulse with a waiting time longer than the cavity decay time to avoid photons in the cavity injected by the readout pulse causing a Stark shift of the qubit resonant frequency.

\subsection{Error probability in feedback control}
\begin{figure}[tb]
 \begin{center}
    \includegraphics[width=16.0cm]{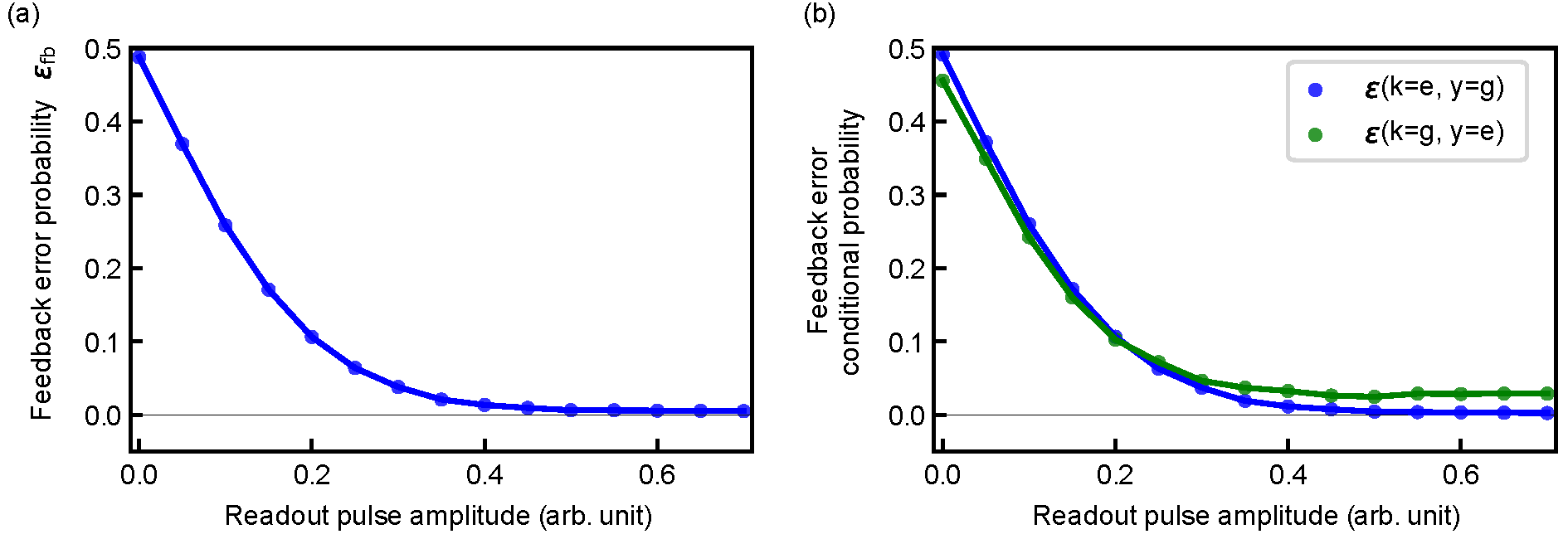}
  \caption{
Feedback error probability vs.\ readout pulse amplitude. 
The effective temperature of the initial state of the qubit in this measurement and the experiment of Fig.~3 is prepared at 0.14~K (the occupation probability of the excited state is 0.097).  
(a)~Feedback error probability $\epsilon_\mathrm{fb}$. (b)~Conditional error probabilities $\epsilon(k=\mathrm{e}\,|\,y=\mathrm{g})$~(blue) and $\epsilon(k=\mathrm{g}\,|\,y=\mathrm{e})$~(green).
} \label{fig:(exp)160508CDK36_651_JEFB_5RO_FBerror_16}
 \end{center}
\end{figure}

Here, we consider the error probability in the feedback control.
Whether or not to inject the $ \pi $-pulse as the feedback control is determined by the variable-strength readout (outcome $ k $) in Fig.~3, while the state of the qubit after the readout is immediately confirmed by the strong measurement (outcome $ y $). 
Events in which $ k $ and $ y $ are not the same are counted as errors of the feedback control.
Let $ n(y=\mathrm{g} ,\, k=\mathrm{e})$ [$ n(y= \mathrm{e} ,\, k=\mathrm{g}) $] denote the number of events, where the $\pi$-pulse is (not) injected erroneously when the qubit state observed as $y$ is the ground state (the first excited state).
Let $n_\mathrm{all}$ denote the total number of the repeated postselected sequences; then the probabilities for these errors can be written as [see Fig.~\ref{fig:(exp)160508CDK36_651_JEFB_5RO_FBerror_16}(a)]
\begin{align}
\epsilon(y=\mathrm{g} ,\, k=\mathrm{e}) := \frac{n(y=\mathrm{g} ,\, k=\mathrm{e})}{n_\mathrm{all}}, \\
\epsilon(y= \mathrm{e} ,\, k=\mathrm{g}):= \frac{n(y= \mathrm{e} ,\, k=\mathrm{g})}{n_\mathrm{all}} .
\label{eq:ebError}
\end{align}
The error probability of the feedback control in Fig.~3 is defined as 
\begin{align}
\epsilon _\mathrm{fb} := 
\epsilon(y=\mathrm{g} ,\, k=\mathrm{e})+\epsilon(y= \mathrm{e} ,\, k=\mathrm{g}).
\end{align}

The feedback operations with these errors are modeled as
\begin{align} 
\hat{I}' :=& \sqrt{1- \epsilon(y=\mathrm{g} ,\, k=\mathrm{e})}\, \hat{I} + \sqrt{\epsilon(y=\mathrm{g} ,\, k=\mathrm{e})}\, \hat{\sigma}_x, \\
\hat{U}_\mathrm{\pi}' :=& \sqrt{\epsilon( y= \mathrm{e} ,\, k=\mathrm{g})}\, \hat{I} + \sqrt{1-\epsilon(y= \mathrm{e} ,\, k=\mathrm{g})}\, \hat{\sigma}_x .
\end{align}
The probability $1 - \lambda_\mathrm{fb}$ that the time-reversed events have their counterparts in the forward process under the feedback control is evaluated from Eq.(\ref{eq:lambda_fb}) and the first line of Eq.(\ref{eq:derivation_of_FT}) as
\begin{align}
1 - \lambda_\mathrm{fb} =& \,
\mathrm{Tr}[p(k=\mathrm{g})\, \langle \mathrm{g} | \hat I'^\dag \hat{\rho}_\mathrm{r} \hat I' | \mathrm{g} \rangle] 
+  \mathrm{Tr}[p(k=\mathrm{e})\, \langle \mathrm{e} | \hat U'^\dag \hat{\rho}_\mathrm{r} \hat U' | \mathrm{e} \rangle] \nonumber \\
=& \, p(k=\mathrm{g})\, [p _\mathrm{can}(\mathrm{g})\,  \{ 1 - \epsilon(y=\mathrm{g} ,\, k=\mathrm{e}) \} + p _\mathrm{can} (\mathrm{e})\, \epsilon( y=\mathrm{g} ,\, k=\mathrm{e})  ] \nonumber \\ 
&+ p(k=\mathrm{e})\, [p _\mathrm{can}(\mathrm{e})\, \epsilon(y= \mathrm{e} ,\, k=\mathrm{g}) + p_\mathrm{can}(\mathrm{g})\, \{ 1 -\epsilon(y= \mathrm{e} ,\, k=\mathrm{g}) \}],
\label{eq:JEFBexp_rhs}
\end{align}
where $p(k=\mathrm{g})$ and $p(k=\mathrm{e})$ are the probabilities of observing the measurement outcomes $k=\mathrm{g}$ and $k=\mathrm{e}$ in the forward process, respectively.
The initial state of the time-reversed process is the canonical state
$\hat{\rho}_\mathrm{r} =p_\mathrm{can}(\mathrm{g}) |\mathrm{g} \rangle \langle \mathrm{g} | + p_\mathrm{can}(\mathrm{e})  | \mathrm{e}  \rangle \langle \mathrm{e} | = \frac{1}{Z} \left( |\mathrm{g} \rangle \langle \mathrm{g} | +  \mathrm{e}^{- \beta \hbar \omega_\mathrm{q}} | \mathrm{e}  \rangle \langle \mathrm{e} | \right)$, where $Z$ is the partition function.

To characterize the origins of the error more precisely,  we define the conditional error probabilities of the feedback control  as
\begin{align}
\epsilon(k=\mathrm{e}\,|\,y=\mathrm{g}):= \frac{\epsilon(y=\mathrm{g} ,\, k=\mathrm{e})}{p(y=\mathrm{g})}, \\
\epsilon(k=\mathrm{g}\,|\,y=\mathrm{e}):=\frac{\epsilon(y=\mathrm{e},\, k=\mathrm{g})}{p(y=\mathrm{e})}.
\end{align}
Note that the probabilities are conditioned on outcome $y$. 
As shown in Fig.~\ref{fig:(exp)160508CDK36_651_JEFB_5RO_FBerror_16}(b), these errors are slightly asymmetric.
There are two reasons for the asymmetry. 
One is the relaxation of the qubit occurring between the two readouts for $k$ and $y$, which is dominant for the strong measurement. 
The other is the small offset in the threshold voltage 
for discriminating the readout signal, which was slightly biased in favour of signalling the excited state. 
In the case of the weak readout pulse amplitude, the error probability $\epsilon(k=\mathrm{e}\,|\,y=\mathrm{g})$ is larger than $\epsilon(k=\mathrm{g}\,|\,y=\mathrm{e})$ due to the offset.

\subsection{Quantum trajectory method}
The quantum trajectory method (Monte Carlo wave function method) is one of the methods to calculate the relaxation of a small quantum system coupled to a heat bath~\cite{Dalibard1992,Molmer1993}．
We calculate the temporal evolution of the quantum system using a state vector (wave function) and a non-Hermitian Hamiltonian.
It is equivalent to the calculation based on the quantum master equation~\cite{Gardiner2004}, but has an advantage that it enables explicit calculation of each trajectory of the system evolution.

The state vector after an infinitesimal time $ \delta t $ is calculated from the state vector $ | \Psi (t) \rangle $ of the qubit at time $ t $ by using the non-Hermitian Hamiltonian
\begin{align}
\hat{H} = \hat{H}_\mathrm{s} - \frac{i \hbar}{2} \sum _{k=0,1} \hat{L}_k^\dagger  \hat{L}_k,
\end{align}
where $\hat{H}_\mathrm{s}$ is the system Hamiltonian and $\hat{L}_k$ is the Lindblad operator.
Here we ignore the effect of dephasing and only consider the relaxation which is relevant to the protocols in the present work.
Thus, we use $\hat{L}_0 = \sqrt{\Gamma _\downarrow } \hat{\sigma}^-$, and $\hat{L}_1 = \sqrt{\Gamma _\uparrow } \hat{\sigma}^+$, where $\Gamma _\uparrow$ and $\Gamma _\downarrow$ are the excitation and relaxation rates of the qubit, respectively, and $\hat{\sigma}^+ = (\hat{\sigma}^-)^\dagger$ is the raising operator of the qubit.
The state vector at time $t+\delta t$ is obtained as
\begin{align}
|\Psi ' (t + \delta t) \rangle &= \exp(- i \hat{H} \delta t /\hbar) |\Psi (t) \rangle \nonumber \\
&\thickapprox \left( 1-\frac{i \hat{H} \delta t}{\hbar} \right) |\Psi (t) \rangle.
\end{align}
Since the above state vector is obtained by applying the non-Hermitian operator, it is not normalized. 
Up to the leading order in $ \delta t $, the norm of the state vector is given as $\langle \Psi ' (t + \delta t) |\Psi ' (t + \delta t) \rangle 
\thickapprox 1 -\delta p$, where we define 
\begin{align}
\delta p &= \delta t \frac{i}{\hbar} \langle \Psi  (t)|\left( \hat{H} - \hat{H}^\dagger  \right) |\Psi (t) \rangle = \sum _{k=0,1} \delta p_k \\
\delta p_k &= \delta t  \langle \Psi  (t)| \hat{L}^\dagger_k \hat{L}_k  |\Psi (t) \rangle \geq 0.
\end{align}
Here, $\delta p = \delta p_0$ when the qubit is in the excited state and $\delta p = \delta p_1$ when it is in the ground state.
Then, to decide whether a quantum leap occurs or not, a random number $\epsilon$ uniformly distributed from $ 0 $ to $ 1 $ is selected at each time step and compared with $\delta p$.

If $ \delta p < \epsilon $, the state evolves to 
\begin{align}
|\Psi (t + \delta t) \rangle = \frac{|\Psi ' (t + \delta t) \rangle }{\sqrt{1- \delta p}}.
\end{align}
On the other hand, if $ \delta p> \epsilon $, the state jumps to a new state vector.
The probability of choosing an individual state vector $\hat{L}_k  |\Psi (t) \rangle$ is $\delta p_k/ \delta p$.
After the normalization, the new state vector is written as 
\begin{align}
|\Psi (t + \delta t) \rangle 
= \frac{ \hat{L}_k |\Psi (t) \rangle }{\sqrt{ \langle \Psi  (t)| \hat{L}_k ^\dagger \hat{L}_k  |\Psi (t) \rangle }}
= \frac{ \hat{L}_k |\Psi (t) \rangle }{\sqrt{\delta p_k / \delta t }}.
\end{align}

We apply the above method in the numerical calculations of the qubit state evolutions and those of the measurement outcomes under the experimental protocols of Fig.~\ref{fig:(schem)plsSeq}.
The outcome of each readout, $\mathrm{g}$ or $\mathrm{e}$, is evaluated from the sign of the time-averaged value of the $z$-component of the qubit state vector, $ \sum _t \langle \Psi (t) | \hat {\sigma} _z | \Psi (t) \rangle / \Delta T $, during the readout pulse with the width $ \Delta T = 500 $~ns. 
The influence of the feedback error is simulated by inverting the outcome $k$ of the readout with the experimentally obtained error probability~(Fig.~\ref{fig:(exp)160508CDK36_651_JEFB_5RO_FBerror_16}).

%%%%%%%%%%%%%%%%%%

\end{document}